\documentclass[onecolumn, amsmath, showpacs,11pt]{revtex4}
\usepackage{graphicx}
\usepackage{dcolumn}
\usepackage{bm}
\usepackage{color}
\begin{document}
\newcommand{\hs}{\hspace*{0.5cm}}
\newcommand{\vs}{\vspace*{0.5cm}}
\newcommand{\be}{\begin{equation}}
\newcommand{\ee}{\end{equation}}
\newcommand{\bea}{\begin{eqnarray}}
\newcommand{\eea}{\end{eqnarray}}
\newcommand{\ben}{\begin{enumerate}}
\newcommand{\een}{\end{enumerate}}
\newcommand{\bde}{\begin{widetext}}
\newcommand{\ede}{\end{widetext}}
\newcommand{\nn}{\nonumber}
\newcommand{\crn}{\nonumber \\}
\newcommand{\Tr}{\mathrm{Tr}}
\newcommand{\non}{\nonumber}
\newcommand{\noi}{\noindent}
\newcommand{\al}{\alpha}
\newcommand{\la}{\lambda}
\newcommand{\bet}{\beta}
\newcommand{\ga}{\gamma}
\newcommand{\va}{\varphi}
\newcommand{\om}{\omega}
\newcommand{\pa}{\partial}
\newcommand{\+}{\dagger}
\newcommand{\fr}{\frac}
\newcommand{\bc}{\begin{center}}
\newcommand{\ec}{\end{center}}
\newcommand{\Ga}{\Gamma}
\newcommand{\de}{\delta}
\newcommand{\De}{\Delta}
\newcommand{\ep}{\epsilon}
\newcommand{\varep}{\varepsilon}
\newcommand{\ka}{\kappa}
\newcommand{\La}{\Lambda}
\newcommand{\si}{\sigma}
\newcommand{\Si}{\Sigma}
\newcommand{\ta}{\tau}
\newcommand{\up}{\upsilon}
\newcommand{\Up}{\Upsilon}
\newcommand{\ze}{\zeta}
\newcommand{\ps}{\psi}
\newcommand{\Ps}{\Psi}
\newcommand{\ph}{\phi}
\newcommand{\vph}{\varphi}
\newcommand{\Ph}{\Phi}
\newcommand{\Om}{\Omega}

\title{Discriminating the minimal 3-3-1 models} 

\author{P. V. Dong}
\email {pvdong@iop.vast.ac.vn} \affiliation{Institute of Physics, Vietnam Academy of Science and Technology, 10 Dao Tan, Ba Dinh, Hanoi, Vietnam}
\author{D. T. Si}
\email {dtsi@grad.iop.vast.ac.vn} \affiliation{Institute of Physics, Vietnam Academy of Science and Technology, 10 Dao Tan, Ba Dinh, Hanoi, Vietnam}
\date{\today}

\begin{abstract}
We show that due to the $\rho$ parameter bound and the Landau pole limit, the reduced 3-3-1 model is unrealistic, while due to the $\rho$ parameter and FCNCs bounds, the simple 3-3-1 model is experimentally unfavored. All such conditions strictly constrain the gauge symmetry breaking scales of the minimal 3-3-1 model with three scalar triplets.        
\end{abstract}


\maketitle

{\it Introduction}: It is well known that the unsolved questions of the standard model, namely the number of fermion generations, uncharacteristic heaviness of the top quark, strong $CP$ problem, electric charge quantization, neutrino masses, and dark matter, can be addressed by the 3-3-1 models \cite{331m,331r,tquark,palp,ecq,neutrino331,dm331}. Moreover, the $B-L$ dynamics and resulting $R$-parity, leptogenesis, and inflaton can also be realized by this kind of the theories \cite{3311}. 

Recently, there have emerged three versions of the minimal 3-3-1 model, the reduced 3-3-1 model \cite{r331}, the simple 3-3-1 model \cite{s331}, and the minimal 3-3-1 model with three scalar triplets \cite{331t}, which provide new theoretical and phenomenological aspects beyond the old ones. In this work, we will show experimentally favored degrees for such theories, simply replied on their $\rho$ parameter, FCNCs and Landau pole. The $Z$ and new $Z'$ gauge boson mixing is also analyzed.      

{\it The minimal 3-3-1 models}: The gauge symmetry is given by $SU(3)_C\otimes SU(3)_L\otimes U(1)_X$ (3-3-1), where the first factor is the color group while the last two are the extension of the electroweak symmetry. The electric charge operator takes the form $Q=T_3-\sqrt{3}T_8+X$, where $Y=-\sqrt{3}T_8+X$ is the weak hypercharge. Here, $T_i (i=1,2,3,...,8)$ and $X$ are the $SU(3)_L$ and $U(1)_X$ charges, respectively (the color charges will be denoted by $t_i$). The fermions can be arranged as $\psi_{aL}=(\nu_{aL},\ e_{aL},\ e_{aR}^c)\sim (1,3,0)$, $Q_{\al L}=(d_{\al L},\ -u_{\al L},\ J_{\al L})\sim (3,3^*, -1/3)$, $Q_{3L}=(u_{3L},\ d_{3L},\ J_{3L})\sim (3,3, 2/3)$, $u_{a R}\sim (3,1,2/3)$, $d_{aR}\sim (3,1,-1/3)$, $J_{\al R}\sim (3,1,-4/3)$, and $J_{3R}\sim (3,1,5/3)$, where $a=1,2,3$ and $\al =1,2$ are generation indices. Note that the values in parentheses present quantum numbers based upon the 3-3-1 symmetries, respectively. 

The minimal 3-3-1 model with three scalar triplets works with the following scalar fields $\eta=(\eta^0_1,\ \eta^-_2,\ \eta^+_3)\sim (1,3,0)$, $\rho=(\rho^+_1,\ \rho^0_2,\ \rho^{++}_3)\sim (1,3,1)$, and $\chi=(\chi^-_1,\ \chi^{--}_2,\ \chi^0_3)\sim (1,3,-1)$. The reduced 3-3-1 model works with ($\rho$, $\chi$) by excluding $\eta$, while the simple 3-3-1 model works with $(\eta,\ \chi)$ by excluding $\rho$. The VEVs of the scalars are given by $\langle \eta\rangle = (u/\sqrt{2},\ 0,\ 0)$, $\langle \rho \rangle = (0,\ v/\sqrt{2},\ 0)$, and $\langle \chi \rangle = (0,\ 0,\ w/\sqrt{2})$. The following calculations generally apply for all the models (for the reduced 3-3-1 model, $u=0$; for the simple 3-3-1 model, $v=0$).       

{\it Gauge boson masses and mixing}: We now derive the mass spectrum of the gauge bosons, which arises from the Lagrangian $\sum_{\Phi=\eta,\rho,\chi} (D_\mu \langle \Phi\rangle )^\dagger (D^\mu \langle \Phi\rangle )$, where the covariant derivative takes the form $D_\mu =\pa_\mu +ig_s t_i G_{i\mu}+ig T_i A_{i \mu }+i g_X X B_\mu$, with the gauge couplings ($g_s,\ g,\ g_X$) and gauge bosons $(G_{i\mu},\ A_{i\mu},\ B_\mu)$, associated with the respective 3-3-1 groups. We have physical charged gauge bosons with respective masses,
\bea && W^\pm\equiv \fr{A_1\mp i A_2}{\sqrt{2}},\hs X^\pm\equiv \fr{A_4\pm i A_5}{\sqrt{2}},\hs Y^{\pm\pm}\equiv \fr{A_6\pm i A_7}{\sqrt{2}},\crn
&& m^2_{W}=\fr{g^2}{4}(u^2+v^2),\hs m^2_X=\fr{g^2}{4}(u^2+w^2),\hs m^2_Y=\fr{g^2}{4}(v^2+w^2).\label{wxymass} \eea To keep consistency with the standard model, we impose $u,v\ll w$. The field $W$ is identical to the standard model charged gauge boson, which implies $v^2_{\mathrm{w}}\equiv u^2+v^2=(246\ \mathrm{GeV})^2$, while $X$ and $Y$ are new gauge bosons with large masses in $w$ scale. 

For the neutral gauge bosons, the photon, $Z$, and new $Z'$ can be identified as \bea            
A&=& s_W A_{3}+ c_W\left(-\sqrt{3}t_W A_{8}+\sqrt{1-3t^2_W}B\right),\crn
Z&=& c_W A_{3}- s_W\left(-\sqrt{3}t_W A_{8}+\sqrt{1-3t^2_W}B\right),\\
Z' &=& \sqrt{1-3t^2_W}A_{8}+\sqrt{3}t_W B,
\eea    
where $s_W=e/g = t/\sqrt{1+4t^2}$, with $t=g_X/g$, is the sine of the Weinberg angle \cite{donglong}. The photon field $A$ is physical ($m_A=0$) and decoupled, whereas $Z$ and $Z'$ mix as given by the mass matrix, 
\bea  \left(\begin{array}{cc}
m^2_Z & m^2_{ZZ'}\\
m^2_{ZZ'} & m^2_{Z'}
\end{array}\right), \eea where 
\bea m^2_Z &=&\fr{g^2}{4c^2_W}(u^2+v^2),\hs m^2_{ZZ'}=\fr{g^2\left[(1-4s^2_W)u^2-(1+2s^2_W)v^2\right]}{4\sqrt{3}c^2_W\sqrt{1-4s^2_W}},\crn 
m^2_{Z'}&=&\fr{g^2\left[(1-4s^2_W)^2u^2+(1+2s^2_W)^2v^2+4c^4_W w^2\right]}{12 c^2_W(1-4s^2_W)}.\eea 
Hence, we obtain two physical neutral gauge bosons (besides the photon),
\be Z_1=c_\varphi Z -s_\varphi Z',\hs Z_2 = s_\varphi Z + c_\varphi Z', \ee with the $Z$-$Z'$ mixing angle,
\be t_{2\varphi}= \fr{2m^2_{ZZ'}}{m^2_{Z'}-m^2_Z}\simeq \fr{\sqrt{3(1-4s^2_W)}}{2c^4_W}\fr{\left[(1-4s^2_W)u^2-(1+2s^2_W)v^2\right]}{w^2},\ee  
and their masses,
 \bea m^2_{Z_1} &=& \fr{1}{2}[m^2_Z+m^2_{Z'}-\sqrt{(m^2_Z-m^2_{Z'})^2+4m^4_{ZZ'}}]\simeq \fr{g^2}{4c^2_W}(u^2+v^2),\\
m^2_{Z_2}&=& \fr{1}{2}[m^2_Z+m^2_{Z'}+\sqrt{(m^2_Z-m^2_{Z'})^2+4m^4_{ZZ'}}]\simeq \fr{g^2c^2_W}{3(1-4s^2_W)}w^2. \eea The approximations for the masses are given at the leading order. Because the mixing angle $\varphi$ is small, we have $Z_1\simeq Z$ and $Z_2\simeq Z'$, which imply that the $Z_1$ is like the standard model $Z$ boson, while $Z_2$ is a new neutral gauge boson with a large mass in $w$ scale. 

{\it $\rho$-parameter}: The experimental $\rho$ parameter (or $\Delta\rho\equiv \rho-1$ used below) that is contributed (or induced) only by the new physics comes from the following sources. The first one is given at the tree-level due to the $Z$-$Z'$ mixing, which can be evaluated as 
\be (\Delta \rho)_{\mathrm{tree}} \equiv \fr{m^2_W}{c^2_W m^2_{Z_1}}-1\simeq \fr{m^4_{ZZ'}}{m^2_Z m^2_{Z'}}\simeq \fr{\left[(1-4s^2_W)u^2-(1+2s^2_W)v^2\right]^2}{4c^4_W v^2_{\mathrm{w}}w^2}.\label{deltarho} \ee
The second one arises from the one-loop contributions of a heavy gauge boson doublet $(X^-,\ Y^{--})$. Note that the other new particles such as the exotic quarks, $Z'$, and new Higgs bosons do not contribute \cite{stu331}. Generalizing the results in \cite{stu331} and using the $X,\ Y$ masses in (\ref{wxymass}), we obtain \bea (\Delta \rho)_{\mathrm{rad}}&=&\fr{3\sqrt{2}G_F}{16\pi^2}\left(m^2_Y+m^2_X-\fr{2m^2_Y m^2_X}{m^2_Y-m^2_X}\ln\fr{m^2_Y}{m^2_X}\right)\crn
&&+\fr{\al}{4\pi s^2_W}\left(\fr{m^2_Y+m^2_X}{m^2_Y-m^2_X}\ln\fr{m^2_Y}{m^2_X}-2+3 t^2_W\ln\fr{m^2_Y}{m^2_X}\right)\crn
&=&\fr{3g^2}{64\pi^2 v^2_{\mathrm{w}}}\left(v^2_{\mathrm{w}}+2w^2-\fr{2(v^2+w^2)(u^2+w^2)}{v^2-u^2}\ln\fr{v^2+w^2}{u^2+w^2}\right)\crn
&&+\fr{g^2}{16\pi^2}\left(\fr{v^2_{\mathrm{w}}+2w^2}{v^2-u^2}\ln\fr{v^2+w^2}{u^2+w^2}-2+3t^2_W\ln\fr{v^2+w^2}{u^2+w^2}\right),\eea where $\sqrt{2}G_F=1/v^2_{\mathrm{w}}$ and $\al=g^2s^2_W/(4\pi)$. Summarizing the above results, we get the $\Delta\rho$ deviation due to the new physics contributions up to one-loop level,  
\be \Delta \rho = (\Delta \rho)_{\mathrm{tree}}+(\Delta \rho)_{\mathrm{rad}}.\label{dongggg} \ee 

Note that $(\Delta\rho)_{\mathrm{rad}}$ can be negative or positive, depending on the sign and magnitude of the $X$ and $Y$ mass splitting, while $(\Delta\rho)_{\mathrm{tree}}$ is always positive. Also, the $(\Delta\rho)_{\mathrm{rad}}$ and $(\Delta\rho)_{\mathrm{tree}}$ are in the same order of $(u/w,v/w)^2$ and can become comparable. On the other hand, it is well-known that the quantum contributions (of the new physics) to $\rho$ depend on only the $X,Y$ mass splitting, that is used to break the vector part of weak $SU(2)$ (see \cite{pdg}). Hence, the multi-loop corrections to $\rho$ are expected to be suppressed by the mass splitting $(m^2_Y-m^2_X)/(m^2_Y+m^2_X)\sim (v^2-u^2)/w^2$ as well as the loop factor $1/(16\pi^2)$, which are only the subleading effects to the leading one-loop result. Our following conclusions based on the one-loop calculations should remain unchanged.                           

{\it New physics constraints}: Because $Z'$ nonuniversally couples to the ordinary quarks, it gives rise to tree-level FCNCs. These processes can be evaluated that are completely identical to those in \cite{s331} and give a bound: $w>3.6\ \mathrm{TeV}$ (see also \cite{others} for other discussions and constraints on the 3-3-1 breaking scale). On the other hand, since $s^2_W=g^2_X/(g^2+4g^2_X)<1/4$, the model encounters a low Landau pole ($\La$), at which $s^2_W(\La)=1/4$ or $g_X(\La)=\infty$, that is roundly $\La=4-5\ \mathrm{TeV}$, depending on the unfixed 3-3-1 breaking scale ($\mu_{331}<\La$) \cite{landau}. Hereafter, $\La=5$ TeV will be taken into account. From the global fit, the $\rho$ parameter is $\rho=1.00040\pm0.00024$, which is $1.7\ \sigma$ above the standard model expectation $\rho=1$ \cite{pdg}.         

\begin{figure}[!h]
\begin{center}
\includegraphics[scale=0.5]{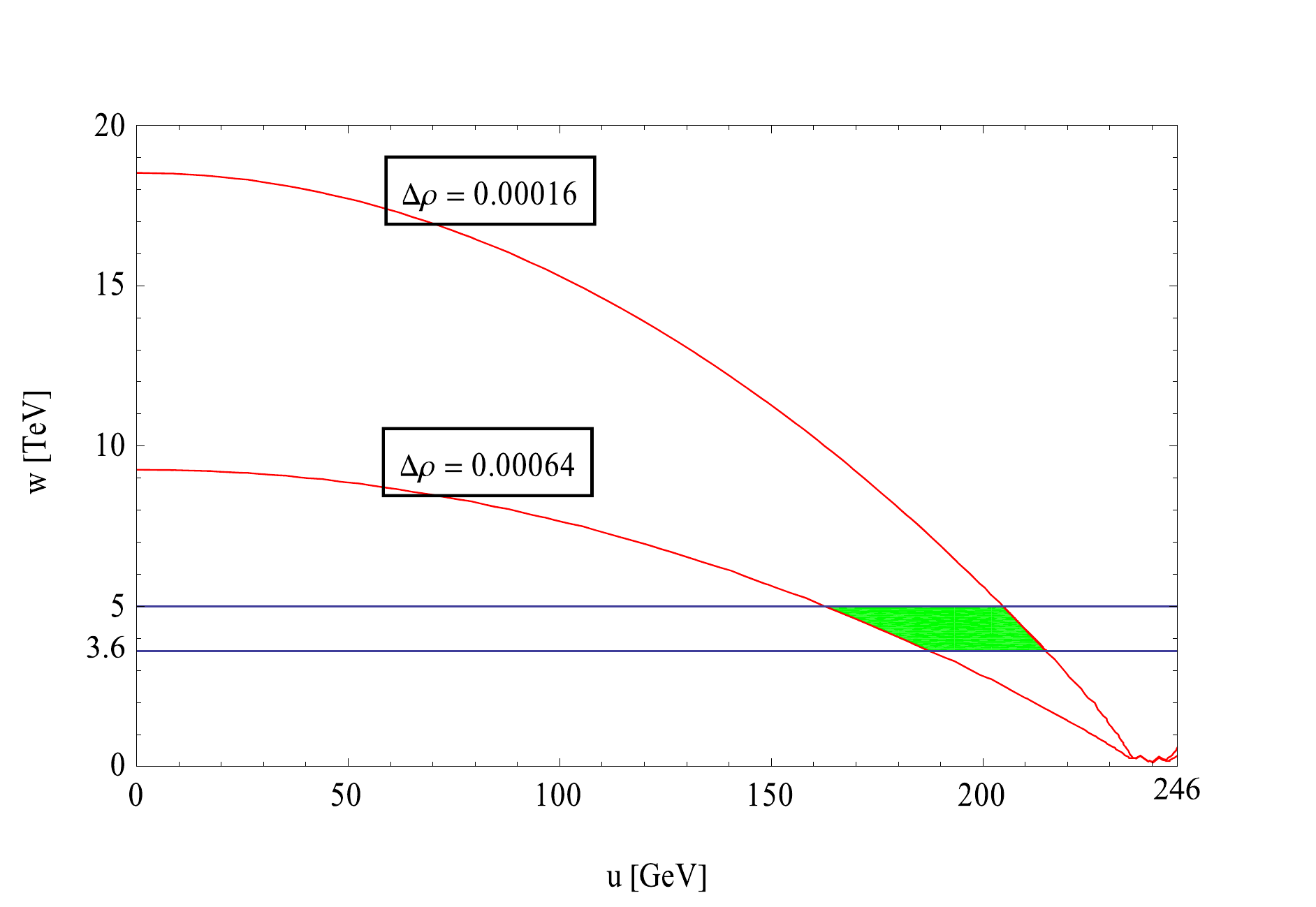}
\caption[]{\label{rho} The $(u,w)$ region that is bounded by $0.00016<\Delta \rho<0.00064$ and $3.6\ \mathrm{TeV} < w<5\ \mathrm{TeV}$. Note that $u$ runs from 0 to $246$ GeV.}
\end{center}
\end{figure}

\begin{figure}[!h]
\begin{center}
\includegraphics[scale=0.5]{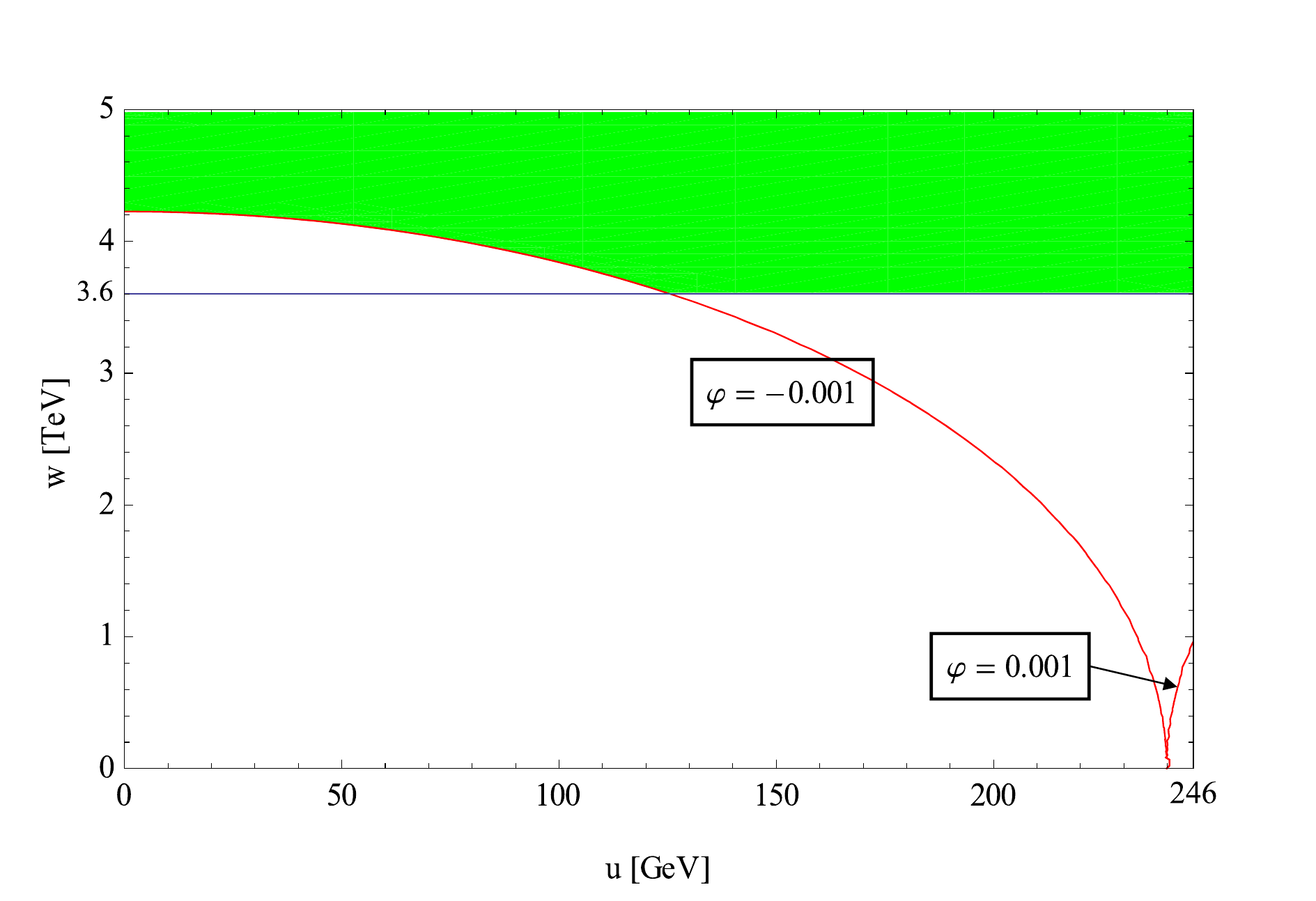}
\caption[]{\label{varphi} The $(u,w)$ region that is bounded by $-0.001<\varphi<0.001$ (the typical limits imposed by the electroweak measurements \cite{pdg}) and $3.6\ \mathrm{TeV}< w<5\ \mathrm{TeV}$. Note also that $u$ runs from 0 to $246$ GeV.}
\end{center}
\end{figure}

Three remarks are in order 
\ben 
\item The reduced 3-3-1 model ($u=0$, $v=v_{\mathrm{w}}$): The deviation $\Delta \rho$ can be approximated as     
\be \Delta \rho \simeq \left(\fr{1+2s^2_W}{2c^2_W}\right)^2\fr{v^2_{\mathrm{w}}}{w^2},\ee which yields $9.243\ \mathrm{TeV}<w<18.487\ \mathrm{TeV}$, provided that $0.00016<\Delta \rho < 0.00064$ and $s^2_W=0.231$ \cite{pdg}. The model is invalid due to the limit of the Landau pole $w<5$ TeV. In other words, due to the Landau pole limit $w<5$ TeV (assumed the model works), we have $\Delta \rho > 0.0022$, which is too large to be consistent with the experimental data \cite{pdg}.        
\item The simple 3-3-1 model ($v=0$, $u=v_{\mathrm{w}}$): The leading order for the $\Delta \rho$ deviation is       
\be \Delta \rho \sim \left[\left(\fr{1-4s^2_W}{2c^2_W}\right)^2+\fr{3\al}{4\pi s^2_W}\left(\fr 1 4 -t^2_W\right)\right]\fr{v^2_{\mathrm{w}}}{w^2},\ee which yields
$w\sim 555 \ \mathrm{GeV}$ (by using the central value $\Delta\rho=0.0004$, $s^2_W=0.231$ and $\al=1/128$ \cite{pdg}). The new physics is well defined below the Landau pole. However, as mentioned the FCNCs constrain $w>3.6\ \mathrm{TeV}$, which opposes the above regime. Thus, the model encounters an experimental discrepancy.          
\item The minimal 3-3-1 model with three scalar triplets: Because of $u^2+v^2=v^2_{\mathrm{w}}$, we can make a contour for $\Delta \rho$ (where $0.00016<\Delta \rho <0.00064$) as a function of only two variables $(u,w)$. The Landau pole limit $w<5$ TeV and the FCNCs bound $w>3.6$ TeV are also imposed. The result is shown in Fig. \ref{rho}. For completeness, the mixing angle $\varphi$ is shown in Fig. \ref{varphi}.      
\een

{\it Conclusion}: The reduced 3-3-1 model should be ruled out because it encounters either a large $\Delta\rho$ deviation or being mathematically inconsistent. The simple 3-3-1 model is experimentally unfavored due to the discrepancy between the FCNCs and $\rho$ parameter bounds. The minimal 3-3-1 model with three scalar triplets is consistent when $3.6\ \mathrm{TeV} < w < 4-5\ \mathrm{TeV}$ and $162.5\ \mathrm{GeV} < u< 215.6\ \mathrm{GeV}$ (or $0.55<v/u<1.14$). In all cases, we can always obtain the corresponding $(u,w)$ values so that the $Z$-$Z'$ mixing angle is small, consistent with the precision data.  

The class of 3-3-1 models with $\beta=\pm\sqrt{3}$ (where $\beta$ determines the embedding of the electric charge operator $Q=T_3+\beta T_8+X$) and basic scalar triplets, which particularly consists of the mentioned ones and the 3-3-1 model with exotic charged leptons \cite{c331}, could be the subject of these constraints. Moreover, although the minimal 3-3-1 model \cite{331m} is not being considered in this work, the FCNCs, $\rho$ parameter, and Landau pole may present the similar bounds besides the others \cite{rstet} when included the additional contributions coming from the scalar sextet.    

However, it is noted that the present constraints might be relaxed because the Landau pole can be lifted up by augmenting the matter content \cite{landau1}. Also, for the 3-3-1 models with $|\beta|<\sqrt{3}$, where $s^2_W=g^2_X/[g^2+(1+\beta^2)g^2_X]<1/(1+\beta^2)$, the Landau pole is lifted. As examples, the 3-3-1 models without exotic charges such as the one with right-handed neutrinos \cite{331r} may be not encountered with a Landau pole up to the Planck scale. Finally, note that the $\rho$ parameter and $Z$-$Z'$ mixing angle also depend on $\beta$, besides the breaking scales $u,v,w$.

\section*{Acknowledgments}

This research is funded by Vietnam National Foundation for Science and Technology Development (NAFOSTED) under grant number 103.01-2013.43.

\end{document}